\documentclass{sag00}

\renewcommand{\arraystretch}{1.5}

\usepackage{multirow}
\usepackage{threeparttable}
\usepackage{graphicx}
\usepackage{xcolor}

\begin{document}
\SetRunningHead{Itoh et al.}{Itoh et al.}

\title{Long term variability of light-pollution in Bisei Town}



%

\author{Ryosuke \textsc{itoh}}
\affil{Bisei Astronomical Observatory, Bisei, Ibara, Okayama, 715-1411, Japan}
\affil{Hiroshima Astrophysical Science Center, Hiroshima University, Higashi-Hiroshima, Hiroshima
739-8526, Japan}
\email{itoh@bao.city.ibara.okayama.jp}
\email{itohryosuke3@gmail.com}
\and
\author{Syota \textsc{Maeno}}
\affil{Bisei Astronomical Observatory, Bisei, Ibara, Okayama, 715-1411, Japan}


%


\maketitle


\begin{abstract}
Bisei town, located in the west part of Japan, is known as a place where the local community protects its beautiful night sky from light pollution through its unique ordinances and the efforts of the local residents. It is also important to monitor in the quantity and quality of light pollution for precise measurement of astronomical observations. The ﬂuorescent lamps in the city were gradually replaced with light emitting diode (LED) lamps. In order to investigate how much light pollution is affecting astronomical observation, we analyzed the archival photometric and spectroscopic data taken by the 101cm telescope that has been installed at Bisei Astronomical Observatory (BAO) since 2006. As a result, we found that there is no signiﬁcant variability in sky brightness in optical bands, but from spectroscopic observation, we observed a blue humps around 4500 \AA originating from LED lights from 2017 to 2023. The brightness of light pollution observed at BAO is not varied but the origin of light has gradually changed from fluorescent lamps toLED lamps.

\end{abstract}

\section{Introduction}
Light pollution originates from the excessive or improperly arranged usage of artiﬁcial light. It is suggested that it affects not only astronomical observation but also ecosystems (e.\ g.,\ \cite{2023NatSR..1317091W, Davies+13, Holker+21} ). \cite{Kyba+17} reported that Earth's artificially lit outdoor area and brightness rate grew by $\sim 2\%$ per year, from 2012 to 2016. The author also claim that only a few countries keep or decrease the sky brightness. For astronomical observation, having a bright sky as a background affects the limiting magnitude that can be observed. This has a signiﬁcant negative impact on the observations of faint, distant galaxies etc. The impact is signiﬁcant not only from a scientiﬁc perspective, but also from an educational one. The experience of observing the stars with the naked eye is one of the ﬁrst step toward scientiﬁc curiosity, but that opportunity is greatly reduced by light pollution. In order to minimize the effects of light pollution, it is crucial to conduct studies on its nature over time. 

Sky brightness has been measured in various ways: by photographing the night sky with a digital camera fitted with a wide-angle lens; by using SKy Quality Meter with Lens (SQM-L) provided by Unihedron; by a CCD camera attached to a telescope, and so on. Each has its advantages and disadvantages (ease/accuracy of measurement, etc...), but in this paper, we use the third method, which accurately measures brightness. When monitoring the variability of light pollution, not only imaging is important but also the spectroscopy of the sky especially since the lights in the city have changed over the past decade from fluorescent lamps to light emitting diodes (LED).

Lights emitted from the city are scattered by the atmosphere and can be observed as light pollution even from a distance of 100 km \citep{Kocifaj+21}.  There are a variety of physical processes that cause diffusion of light pollution; Mie or Rayleigh scattering, depending on he size/shape of aerosols/particles, and not only scattering but also some aerosols/particles could be absorbers. However, the theoretical prediction of light pollution is difficult since there are several types of light sources, as well as the presence of ﬂoating clouds and atmospheric inhomogeneity \citep{Kocifaj+23}. There are several studies of light pollution, especially by photometric observation (e.g., \cite{onoma+09}), but spectroscopic observations alone are not enough in Japan, one reason for this may be the lack of spectrometers among amateurs\citep{ie+91,Kato+20}. As the ﬁrst step toward unraveling this mechanism, it is crucial to measure the spectrum of light.

Bisei Astronomical Observatory (BAO) is a public observatory that opened on 7 July 1993, and is located in Bisei Town, Ibara City, Okayama Prefecture in Japan. The town is also known as the first local government in Japan to enact a light pollution prevention ordinance in 1989. The observatory has an optical telescope with a 101cm aperture, and is open to the public on weekends and at night, attracting many visitors. The telescope is also open to amateurs on weekend evenings. BAO has been using the 101cm telescope for 30 years since its opening, and has accumulated a large amount of imaging observation data. By reanalyzing these data, it is possible to accurately measure the sky brightness and spectrum from past to present.

\section{Observation}
The main telescope of BAO is a 101cm optical telescope mounted on an equatorial mount. The telescope has four types of focus; the main Cassegrain focus is for visual observation, the north-folded Cassegrain focus has an optical imaging monochromatic CCD for photometry, the west-folded Cassegrain focus is where a (commercial) digital camera can be mounted, and the south Cassegrain focus has an optical spectrograph. Our observatory has observed a wide variety of objects; super novae, novae, active galactic nuclei, electromagnetic counterparts to gravitational wave events, variable stars, minor planets, artificial satellites and so on. We have a vast amount of archived data, most of which are available for the measurement of sky brightness. From these data, we are able to carry out the long-term monitoring of light pollution.

\subsection{Photometry}
Photometric observation of the sky was performed with a monochromatic CCD camera (SBIG STL-1001E) fitted to the 101cm telescope from 2008 to 2021. We selected data with the following conditions: taken (1) with object altitude over 60 degrees; (2) with {\it B, V} or {\it R}$_{\rm{C}}$-band ﬁlter; (3) when Moon altitude is less than 0 degrees; (4) when Sun altitude is less than -22 degrees (end/before the astronomical twilight); and (5) when the atmospheric transparency is more than 40\% (typical atmospheric transparency is $\sim$60\% at BAO).  
The data reduction involved standard CCD photometry procedures - dark current subtraction, flat fielding and aperture photometry. We calculated the zero-magnitude for each frame with the stars cataloged in UCAC-4 within the field of view. We used the AB magnitude system for the flux calibration. For the calculation of sky brightness, we calculate the median from the values of all pixels in the FITS images, then we converted the count value to the unit of magnitude/arcsec$^2$ using zero-magnitude and pixel-scale for each image. We did not apply to any star mask but confirmed that even in large clusters such as M13, we found that whether or not a mask is applied has little effect on the results (less than 1\%). We also confirmed that the standard deviation for the difference between our measurement of the star field and the cataloged value in UCAC-4 was $\Delta V \sim 0.18$ [mag] for the CCD+101cm telescope. This value was added to the photometric errors. The atmospheric transparency are calculated by the ratio of observed brightness of stars used in the flux calibration and 
theoretically predicted photon amount from telescope optics. It also contains device efficiency (quantum efficiency of CCD, reflectance of Mirrors and so on) but the main contribution of variability for the value is atmospheric condition.

\subsection{Spectroscopy}
The instrument has two different gratings ($n = 300, 1800 $ [/mm]) thus two spectral-resolutions can be selected, one  low-resolution ($R \sim 1,500, \lambda = 4000 \sim 8000$ \AA) and the other middle-resolution ($R \sim 15,000, \lambda \sim 300$ \AA width). We selected the data from our archival data taken with long-slit low-resolution mode in order to investigate the variability in the wide wavelength range. The data was selected based on the following criteria: (1) Altitude of the Sun is less than -22 degrees (end/before the astronomical twilight); (2). Altitude of the Moon is less than 0 degrees; (3) Altitude of the target star is more than 45 degrees. In addition to the above, we selected data which had high S/N that defined as the ratio of continuum flux between 4800 and 5200 \AA and its variance, and low count ratio of sky brightness. Finally, we obtained 18 spectroscopic data from October 2006 to July 2023.

The data reduction was performed under the standard procedure of CCD spectroscopy; bias subtraction, flat-fielding, cutting out the sky region, conversion to one-dimensional spectrum, wavelength calibration and flux calibration. We selected a sky region which did not contain the starlight of spectroscopic data (regions at least 30 arcsec away from star) with a 50-pixel ($\sim 40$ arcsec) width against the direction of the dispersion axis, which determined from the limitation that no distortion or rotation of the emission lines caused by the optical system. Then we took the median value for the spatial direction axis and obtained one-dimensional data. The calibration of wavelength was performed with atmospheric emission lines (typically 10 lines) for each frame. The flux was calibrated using observations of a spectrophotometric standard star. Note that absolute flux were not reliable since collected spectrum are taken by different observers with different method (altitude, time interval between observed frame and standard star, weather condition and so on). We used two different CCD; ANDOR DU-440BV (2006-2021) and DU-940P-BV (2021-2023) for the spectroscopy with the same optics, and also conﬁrmed that the difference of the wavelength sensitivity was quite small and therefore negligible  (difference is less than $\pm2\%$ between 4000 and 7000\AA).

\begin{table}
  \caption{Spectroscopic observation}\label{tab:specobs}
  \begin{center}
  \begin{threeparttable}[h]

    \begin{tabular}{ccrrr}
      \hline
      Year  & Date (UTC) & Altitude\tnote{a} & Azimuth\tnote{b}  & Sun Alt.\tnote{c} \\
      \hline\hline
2006 & 10/29 18:09 & 46.6 & 63.2 & -40.4 \\
2007 & 02/21 16:50 & 54.2 & 175.3 & -58.2 \\
2008 & 04/05 16:14 & 60.1 & 237.4 & -46.1 \\
2009 & 08/31 18:35 & 45.7 & 149.5 & -24.4 \\
2010 & 08/17 14:24 & 70.5 & 84.0 & -40.8 \\
2011 & 12/27 16:40 & 51.5 & 15.9 & -66.6 \\
2012 & 08/17 14:49 & 76.1 & 86.2 & -41.9 \\
2013 & 06/16 15:50 & 58.4 & 157.3 & -31.0 \\
2014 & 01/06 19:05 & 66.8 & 191.1 & -38.0 \\
2015 & 08/22 16:31 & 53.2 & 78.6 & -39.7 \\
2016 & 03/05 14:36 & 56.3 & 37.0 & -59.6 \\
2017 & 11/11 14:04 & 76.8 & 54.5 & -70.1 \\
2018 & 05/19 17:08 & 55.3 & 75.9 & -27.7 \\
2019 & 03/08 15:43 & 74.6 & 124.8 & -59.5 \\
2020 & 11/21 18:11 & 51.4 & 182.7 & -43.8 \\
2021 & 01/17 13:50 & 64.4 & 300.5 & -66.5 \\
2022 & 10/23 14:00 & 66.8 & 336.2 & -64.3 \\
2023 & 07/10 13:12 & 87.7 & 338.6 & -26.5 \\
\hline
    \end{tabular}
        \begin{tablenotes}
    \item[a] telescope altitude [deg],
    \item[b] telescope azimuth angle from the North to East [deg],
    \item[c] Sun altitude [deg].
    \end{tablenotes}
  \end{threeparttable}
  \end{center}
\end{table}

\section{Results}
\label{sec:res}

\subsection{Photometry}

\begin{figure}
  \includegraphics[bb=0 0 432 288,width=8.5cm]{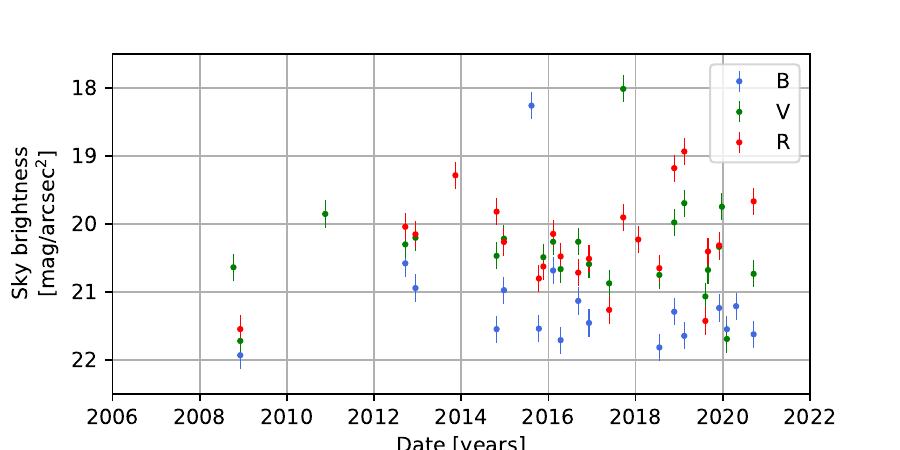}
  \caption{Variability of sky brightness in {\it B, V} and $R_{\rm{C}}$ -bands from 2008 to 2021 taken at BAO. The data represents the darkest sky brightness for that one night.} 
  \label{fig:lc_all}
\end{figure}

\begin{figure}
  \includegraphics[bb=0 0 432 288,width=8.5cm]{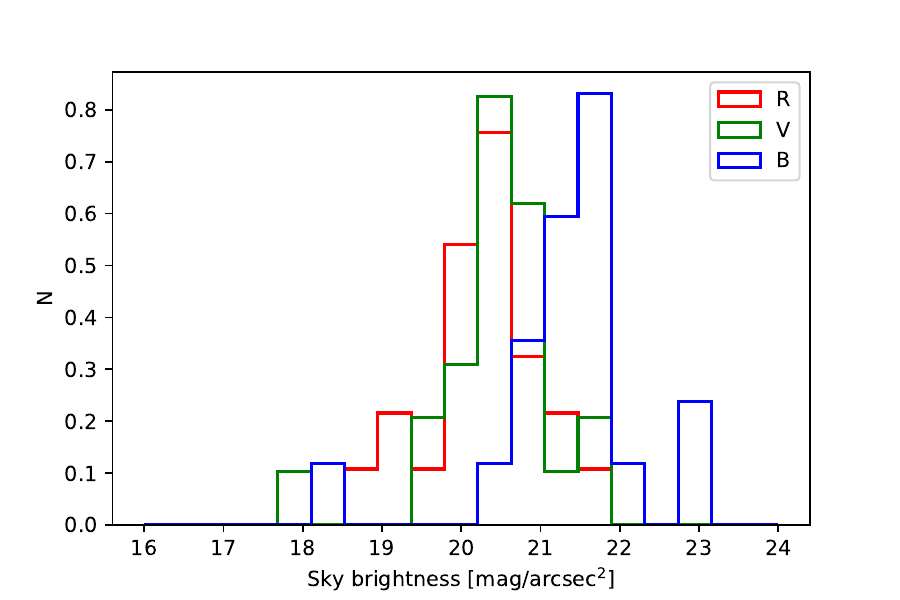}
  \caption{Normarized histogram of sky brightness in {\it B, V} and $R_{\rm{C}}$ -bands from 2008 to 2021 taken at BAO.} 
  \label{fig:hist_all}
\end{figure}

Figure \ref{fig:lc_all} shows the variability of sky brightness in {\it B, V} and {\it R}$_{\rm{C}}$-band from 2008 to 2021 and Figure \ref{fig:hist_all} shows the distribution of sky brightness for each band.  The blue, green, and red data points indicate {\it B, V} and {\it R}$_{\rm{C}}$-band sky brightness, respectively.  The median values of sky brightness are $B \sim 20.9, V \sim 20.3$ and $R_{\rm{C}} \sim 19.8$ [mag/arcsec$^2$] and standard deviation of sky brightness are $B \sim 0.5, V \sim 0.5$ and $R_{\rm{C}} \sim 0.6$. 

We examined the data for variability in the distribution of sky brightness after every three year period, and found that no significant variability was detected in any year for all B, V and  Rc-bands  from  2008 to 2021 (Kolmogorov-Smirnov test, C.L. = 95\%).  The darkest sky brightness levels were {\it B} = 22.0,{\it V} = 22.0 and {\it R}$_{\rm{C}}$ = 21.5 [mag/arcsec$^2$]. The sky brightness darker than 21.0 mag/arcsec$^2$ in the {\it V}-band accounts for about 5\% of the total number of moonless nights observed.

\subsection{Spectroscopy}

\begin{figure}
  \includegraphics[bb=0 0 494 592,width=8.5cm]{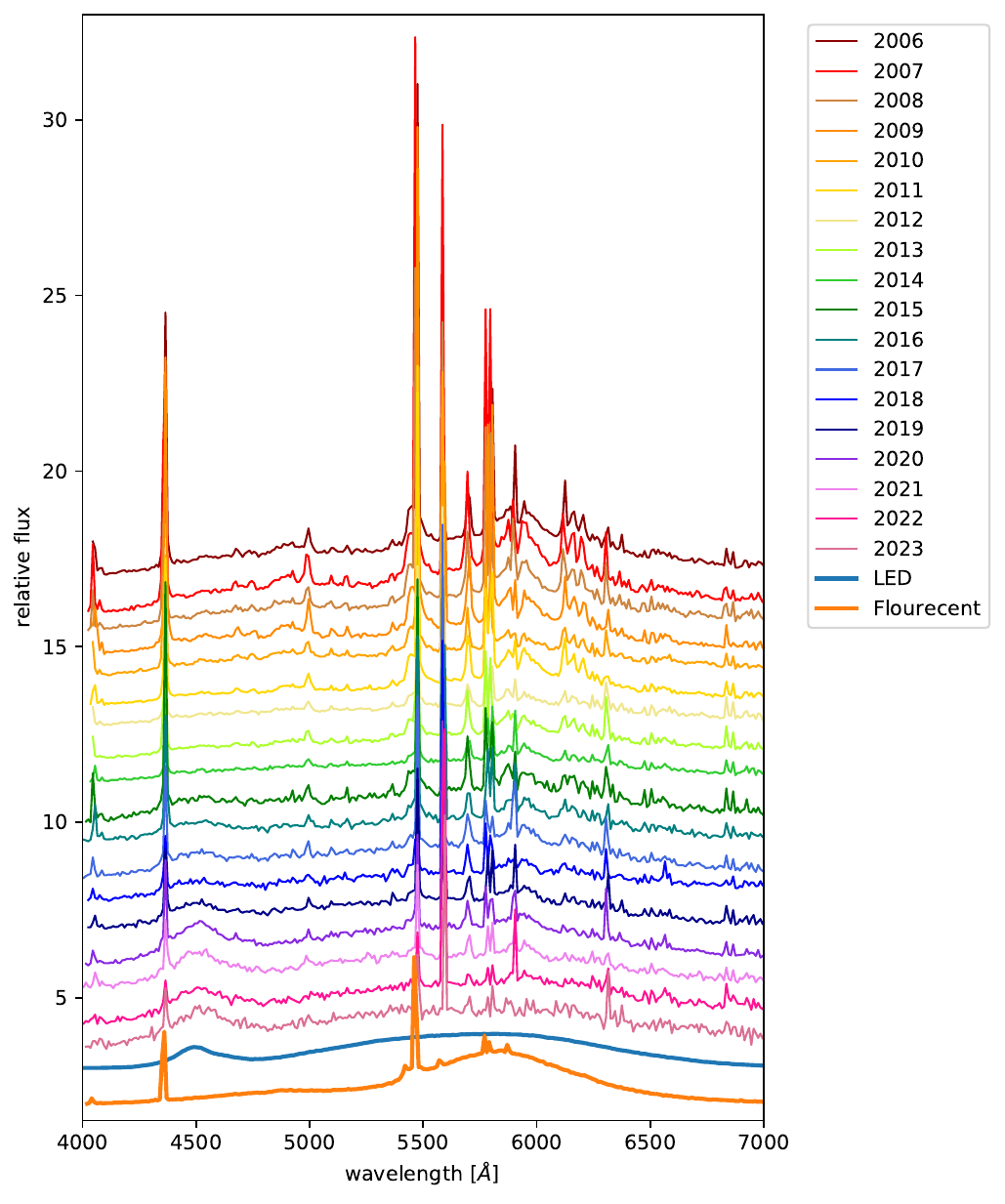}
  \caption{ Sky spectra of BAO from 2006 to 2023. Different colors represent different years. Spectra of LED and fluorescent lamps taken with same instruments are also shown for comparison (bottom two spectra).}
  \label{fig:spec_all}
\end{figure}

\begin{figure}
  \includegraphics[bb=0 0 644 340,width=8.5cm]{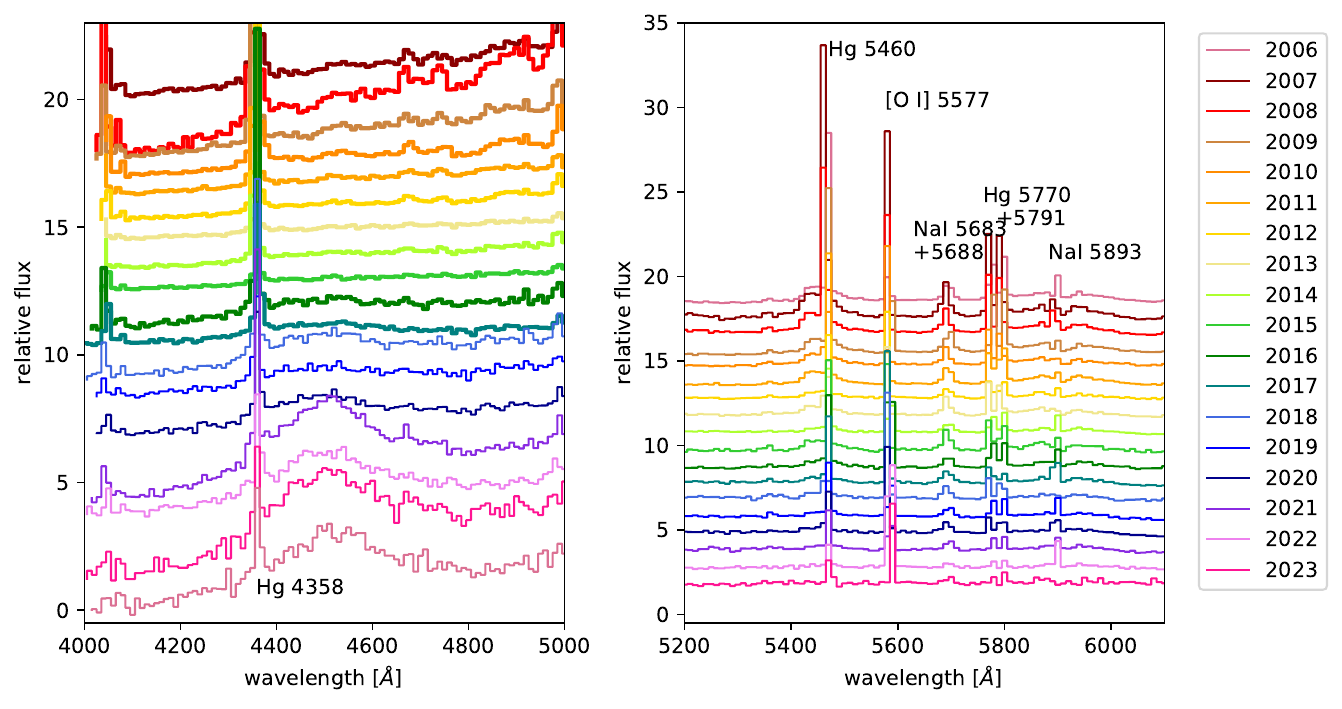}
  \caption{Close up view of spectra around 4500 \AA (left panel) and 5700 \AA (right panel). }
  \label{fig:spec_zoom}
\end{figure}

Figure \ref{fig:spec_all} shows all spectra of the sky from 2006 to 2023. Spectra of the LED and ﬂuorescent lamps installed inside the dome was also taken with same apparatus and is shown for comparison. For the sake of visibility, the ﬂux levels were shifted. Figure \ref{fig:spec_zoom} shows the close-up views of spectra  around 4500 \AA and 5700 \AA, which corresponds to the LED bright part and lines from the ﬂuorescent (Hg) lamps and sodium vapor (Na) lamps.
For the spectrum taken in 2006 (top brown line), strong Hg (4358, 5460, 5770 and 5791 \AA), Na (5683 + 5688 \AA),
is seen and no hump structure around 4500 \AA is visible. With the passing of time, a hump structure is seen around 4500 \AA, and the Hg and Na line seems to decrease. Strong lines of Y(P, V)O$_4$:Eu$^{3+}$ (around 6200 \AA, \citep{Kocifaj+23b}), also originating from the ﬂuorescent lamps were also visible in early observation and decreased in late observation. The characteristic hump around 4500 \AA is also visible in the spectra of the LED light (bottom blue line in Fig \ref{fig:spec_all}) and therefore, the origin of this hump is thought to be due to the LED lamp.

\begin{figure}
  \includegraphics[bb=0 0 440 390,width=8.5cm]{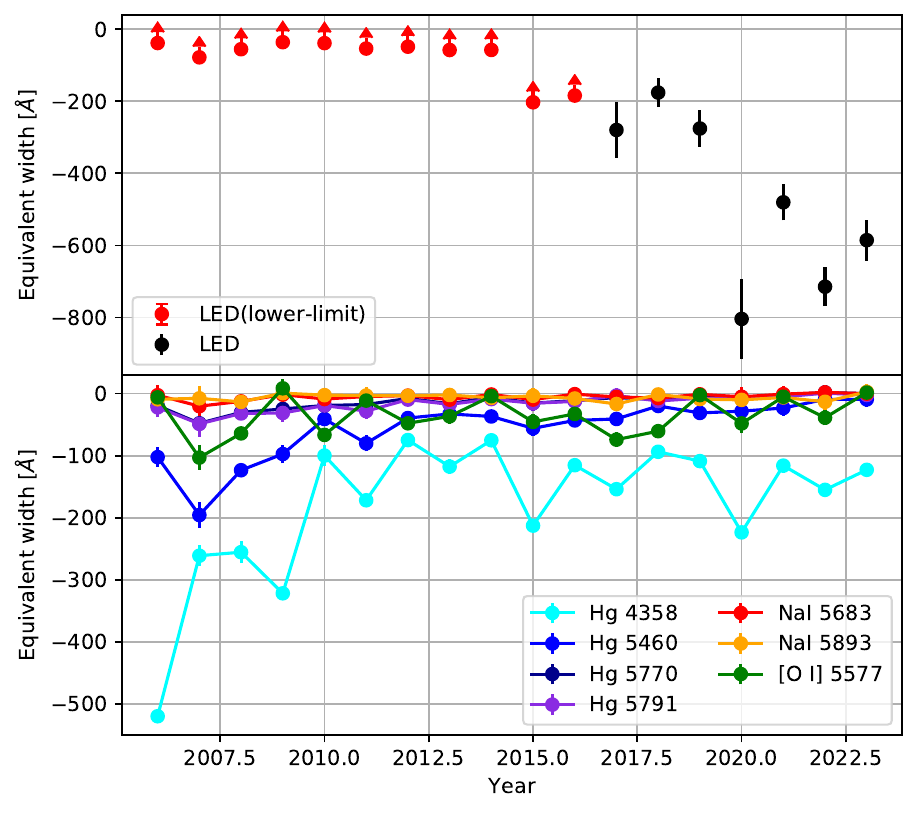}
  \caption{Time variability of equivalent width for LED (upper panel), Hg, Na and [O I] (bottom panel) line from 2006 to 2023.}
  \label{fig:EW}
\end{figure}

In order to investigate the transition of light sources, we derived the equivalent width (EW) of the blue hump around 4500 \AA, Hg, Na and [O I] lines for each spectrum and Figure \ref{fig:EW} shows the time variability of these EWs. 
In order to avoid the contamination of the Hg 4358 line for LED hump, we ﬁtted the $4000 \sim 5000$ \AA (except $4300 \sim 4400$ \AA) spectra with a Gaussian function and linear straight line.EWs and errors of LED hump were calculated based on the integration of Gaussian derived by spectral fitting and propagation of fitting error (1-sigma).  If the difference of chi-square value between linear + Gaussian fitting and linear fitting is less than 3-sigma, we marked the data as non-detection and calculated the lower limit with searching the parameter of Gaussian which makes 3-sigma difference of chi-square. EWs for the other lines were derived by the simple integration of spectra around the target line (width of 30 \AA which covers $\sim 3$-sigma region of the lines). No signiﬁcant (3-sigma) LED hump was detected before 2017. After 2017, an LED hump was detected and became more pronounced with each passing year.

In contrast, the lines originating in the fluorescent lamps (Hg  4358, 5460, 5770 and 5791 \AA) became weaker in later years. 
There is a time lag between the beginnig of decreasing of Hg line and appearance of LED hump. Since this hump is widely spread compared with Hg lines, therefore it is difficult to detect at early phase .

\section{Discussion}

From section \ref{sec:res}, we found that there is no significant variability of the sky brightness in {\it B, V}, and {\it R}$_{\rm{C}}$-bands from the photometric observations, but found the appearance of LED hump structure around 4500 \AA in the spectroscopic observation from 2017 highly significant (this wavelength is corresponds to the {\it B}-band in photometric observation). This appears to be an odd result since the LED hump component should appear in the variability of {\it B}-band sky brightness. We also tested the difference of the sky brightness distribution of the {\it B}-band from the photometric data taken before 2017 and after 2017. The result was that the distribution was quite similar and we found there was no signiﬁcant difference between the data set (Kolmogorov-Smirnov test, 95\% C.L.). Assuming the same baseline light intensity in {\it B}-band, the estimated contribution from the LEDs is about 40\% (corresponding to 0.4 mag) brighter, even when we take into account the reduction of the Hg 4358 line coming from ﬂuorescent lamps. This is clearly inconsistent with the photometric results in B-band and actually implies that a decrease in the baseline flux of 4000-5000 \AA (there is a continuum ﬂux of ﬂuorescent lamps) and the increase of hump structures are occurring simultaneously. As a result, no significant variability in sky brightness can be seen.
Note that absolute flux calibration for spectroscopic observation  is very difficult and our results are shifted to make it easy to see.

In Bisei town, the replacement of fluorescent lamps with LED lamps for the town's street lighting was started around 2011. In 2021, all of the streetlights and outside lamps installed at public facilities were replaced with  light-pollution-preventative LED lamps, which have a light source with a color temperature of 3000K or less, and severely limit upward luminous flux. That being said, white LED lamps which include the blue hump around 4500 \AA are still being used and the light from neighboring cities is very strong. There are three cities with populations over 450,000 within a 50 km radius around Bisei Town; Fukuyama city (450,000 peoples, $\sim$ 30 km), Kurashiki city (470,000 peoples, $\sim$ 25 km) and Okayama city (720,000 peoples, $\sim$ 35 km). It is presumed that light pollution originating from these cities can be observed from the Bisei town.  Quantitative evaluation of how much light from these cities and Bisei town contributes to light pollution observed at BAO by comparison with theoretical  atmospheric scattering models is a future work. In order to comparison with theoretical light pollution models, both observation of absolute brightness by photometric observation and contribution ratio of light source (ﬂuorescent and LED) by spectroscopic observation. And now is a transitional period from fluorescent lights to LEDs, it is important to continue both photometric and spectroscopic observations to investigate the effect of the variation in light source on light pollution.

Observed sky brightness of $V \sim 20.5$ mag/arcsec$^2$ corresponds to class 4 (rural/suburban transition) of the Bortle dark-sky scale \citep{Bortle+01}, and it is consistent with the value calculated with the artificial night sky brightness at Bisei, reported in \cite{Cinzano+01}. Therefore we could say that the night sky in Bisei town is not outstandingly dark, and the sky is affected by light pollution. Since it is difficult to separate the effect of light pollution from other cities with the data used in this paper, we could not investigate how Bisei town’s efforts (replacement of city light, light pollution prevention ordinance and so on) are helping to prevent light pollution. But as mention in above, comparison with theoretical atmospheric scattering models, especially the observations with low altitude and toward the direction of other urban cities might be the key to separate the origin of light pollution.

\section*{Acknowledgement}

We are grateful to the users of 101cm telescope at BAO who provided the sky brightness data. We thank Nobuko Sakagawa for digitization of vast paper observation logs. We also thank anonymous reviewer and Kazuya Ayani fruitful discussions.



\end{document}